\documentclass[jap, reprint, showpacs, superscriptaddress]{revtex4-1} 
\usepackage{graphicx}
\usepackage{amssymb} 
\usepackage{amsmath}
\usepackage{epstopdf}
\usepackage{color}
\usepackage{ulem}

\begin{document} 

\title{Influence of excitonic effects on luminescence quantum yield in silicon}

\author{A.V.~Sachenko}
\affiliation{V. Lashkaryov Institute of Semiconductor Physics, NAS of Ukraine, 41 prospect Nauky, 03028 Kyiv, Ukraine }
\author{V.P.~Kostylyov}
\affiliation{V. Lashkaryov Institute of Semiconductor Physics, NAS of Ukraine, 41 prospect Nauky, 03028 Kyiv, Ukraine }
\author{V.M.~Vlasyuk}
\affiliation{V. Lashkaryov Institute of Semiconductor Physics, NAS of Ukraine, 41 prospect Nauky, 03028 Kyiv, Ukraine }
\author{ I.O.~Sokolovskyi}
\affiliation{V. Lashkaryov Institute of Semiconductor Physics, NAS of Ukraine, 41 prospect Nauky, 03028 Kyiv, Ukraine }
\affiliation{Department of Physics and Physical Oceanography, Memorial University of Newfoundland, St. John's, NL, A1B 3X7  Canada}
\author{ M.~Evstigneev}
\email[Corresponding author: ]{mevstigneev@mun.ca}
\affiliation{Department of Physics and Physical Oceanography, Memorial University of Newfoundland, St. John's, NL, A1B 3X7  Canada}

\begin{abstract}
Nonradiative exciton lifetime in silicon is determined by comparison of the experimental and theoretical curves of bulk minority charge carriers lifetime on doping and excitation levels. This value is used to analyze the influence of excitonic effects on internal luminescence quantum yield at room temperature, taking into account both nonradiative and radiative exciton lifetimes. A range of Shockley-Hall-Reed lifetimes is found, where excitonic effects lead to an increase of internal luminescence quantum yield.

Keywords: silicon, excitons, luminescence, quantum yield, recombination

\end{abstract}


\maketitle 
\section{Introduction}

Excitonic effects in both radiative and nonradiative recombination play an important role in photo- and electroluminescence processes, whose quantum yield is about 10\% and 1\%, respectively \cite{Green01, Ng01}. Their influence on the photoelectric processes in silicon-based devices, such as solar cells, was studied in a number of publications \cite{Kane93, Corkish93, Green98, Gorban00}. 

Investigations of bulk lifetime in silicon as a function of doping and excitation levels were initiated by Hangleiter \cite{Hangleiter88}, who considered the spatial correlation of two electrons and a hole (or two holes and an electron), which exists in the presence of an exciton. 
This correlation leads to an  increased probability of Auger recombination, in which the energy released is transferred to another charge carrier. It can proceed both via the interband and the deep impurity mechanisms.

Excitons' effect on electoluminescence in silicon were considered in \cite{Sachenko06}. However, the effect of nonradiative exciton recombination in n-type silicon was overestimated in that work.

Here, a more accurate value of exciton nonradiative lifetime is obtained by means of comparison between the theoretical and experimental dependences of the bulk lifetime on doping and excitation levels in silicon. In this comparison, improved theoretical expressions for the interband Auger recombination rate from Ref.~\onlinecite{Richter12} are used. The so obtained lifetime served as an input parameter to analyze the contribution of excitonic effects to the luminescence internal quantum yield in silicon. Both the positive effect of radiative exciton recombination and the negative effect of nonradiative recombination are taken into account. Quantum yield dependence on doping and excitation levels, as well as the influence of surface recombination on quantum yield at room temperature are analyzed. It is shown that, at Shockley-Hall-Reed lifetime exceeding 1~ms, radiative exciton recombination dominates; in the opposite case, the negative, i.e., nonradiative effect takes over.

\section{Nonradiative exciton lifetime in silicon}
As shown in Ref.~\onlinecite{Sachenko00} based on the results obtained by Hangleiter \cite{Hangleiter88}, the presence of the exciton subsystem in silicon at sufficiently high doping and excitation levels leads to an onset of the nonradiative exciton recombination channel via Auger mechanism on deep impurities. Its characteristic time is
\begin{equation}
\tau_{nr} = \tau_{SRH} n_x/n\ ,
\label{1}
\end{equation}
where $\tau_{SRH}$ is Shockley-Reed-Hall lifetime, $n = n_0 + \Delta n$ is the density of electron-hole pairs, consisting of the equilibrium, $n_0$, and excess, $\Delta n$, contributions, and $n_x$ is the parameter of the theory. In the case considered here, $n_0$ practically coincides with the doping level. On the other hand, the non-radiative recombination rate can be written as \cite{Sachenko00}
\begin{equation}
\frac{1}{\tau_{nr}} = \frac{n}{n^*}n_L G_t N_t\ ,
\label{1a}
\end{equation}
where $N_t$ is the density of recombination (trap) centers,  $G_t$ the probability of Auger recombination on an impurity,
\begin{equation}
n_L = \frac{3}{4\pi a_B^3},\ n^* = \frac{N_c N_v}{N_x} e^{-E_x/k_BT}\ ,
\label{2}
\end{equation}
$a_B$ being the exciton Bohr radius,  $E_x$ the binding energy of the exciton ground state, and $N_c$, $N_v$, and $N_x$ the effective densities of states in the conduction, valence, and exciton bands in silicon. Combination of Eqs.~(\ref{1}) and (\ref{1a}), taking into account that $\tau_{SRH}^{-1} = C_tN_t$, where $C_t$ is the capture coefficient of a minority carrier by a deep impurity,  gives
\begin{equation}
n_x = \frac{C_t}{G_t}\frac{n^*}{n_L}\ .
\end{equation}
Taking $E_x = 14.7$\,meV, $a_B = 4.2$\,nm, $N_c = 3.12\cdot 10^{19}$\,cm$^{-3}$, $N_v = 2.98\cdot 10^{19}$\,cm$^{-3}$, $N_x = 6.23\cdot 10^{20}$\,cm$^{-3}$, $T = 300$\,K, and assuming that the electron density is well below the Mott transition threshold (about $5\cdot 10^{17}$\,cm$^{-3}$), we obtain $n^* \approx 8\cdot 10^{17}$\,cm$^{-3}$ and $n_L \approx 3.3\cdot 10^{18}$\,cm$^{-3}$ \cite{Kane93}.

As this estimate shows, the ratio $n^*/n_L$ in silicon is of the order of unity. This means that $n_x$ is of the same order of magnitude as the ratio of the capture coefficient by a deep impurity level to the probability of impurity-assisted Auger recombination.

\begin{figure}[t!] 
\includegraphics[scale=0.33]{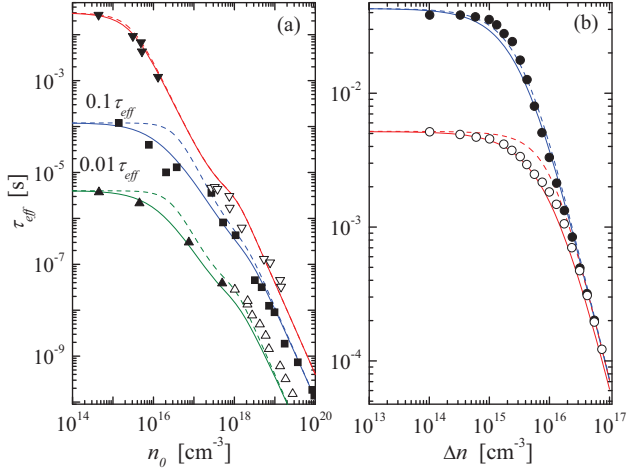}
\caption{Effective bulk lifetime, $\tau_{eff}$, in n- and p-type silicon vs. (a) dopant density and (b) excitation level, obtained experimentally (symbols) and theoretically (solid lines). The theoretical curves were built without (dashed) and with (solid) exciton recombination taken into account. Experimental data in panel (a) are taken from Ref.~\onlinecite{Richter12} (filled down triangles), Ref.~\onlinecite{Fossum76} (filled up triangles), Ref.~\onlinecite{Hacker94} (filled squares), Ref.~\onlinecite{Beck73} (open down triangles), and Ref.~\onlinecite{Passari83} (open up triangles). For the ease of comparison, the curves obtained for the smaller values of $\tau_{max}$ are shifted down on the logarithmic scale. In panel (b), the experimental points are from Ref.~\onlinecite{Yablonovitch86} for $\langle100\rangle$ (filled circles) and $\langle111\rangle$ (open circles) crystallographic orientation of Si.}
\label{fig1}
\end{figure} 

In the work \onlinecite{Sachenko00}, the value of $n_x \approx 3.7\cdot 10^{15}$\,cm$^{-3}$ was determined by analyzing the experimental data for the inverse effective lifetime, $\tau_{eff}^{-1}$ vs. doping level in silicon, published in the literature. There are reasons to believe that this figure underestimates the true value of $n_x$, because, as will be shown below, it does not agree well with the experimental results published in those works, where the surface recombination rate was minimized to such a degree that its influence on $\tau_{eff}$ was negligible. In this case, the effective recombination rate is determined by the Shockley-Hall-Reed recombination, nonradiative exciton recombination, radiative band-to-band and exciton recombination, and interband Auger recombination rates:
\begin{equation}
\tau_{eff}^{-1} = \frac{1 + (n/n_x)}{\tau_{SRH}} + \frac{1}{\tau_{r1}} + \frac{1}{\tau_{r2}} + \frac{1}{\tau_{Auger}}\ .
\label{3}
\end{equation}
Here, $\tau_{r1} = (A_{pn}n)^{-1}$ is the interband recombination lifetime with the respective parameter $A_{pn}$, $\tau_{r2} = (A_xn)^{-1}$ is the radiative exciton recombination lifetime with the parameter $A_x$, see Ref.~\onlinecite{Sachenko06}, and $\tau_{Auger}$ is the interband Auger recombination lifetime given by an empirical expression (18) from Ref.~\onlinecite{Richter12}.

\begin{figure}[t!] 
\includegraphics[scale=0.28]{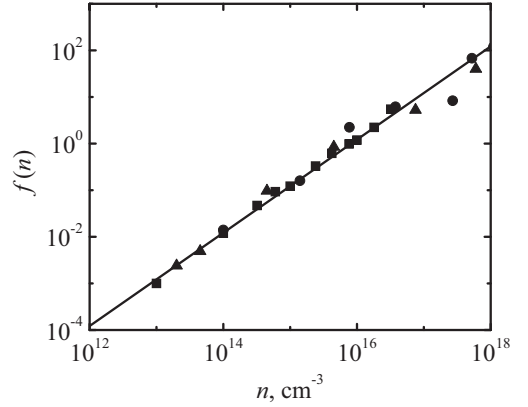}
\caption{Theoretical and experimental dependence of $(\tau_{SRH}/\tau_{eff}^*) - 1$ on the electron density. Symbols: Ref.~\onlinecite{Yablonovitch86} (squares), Ref.~\onlinecite{Fossum76} (triangles), and Ref.~\onlinecite{Hacker94} (circles). Solid line: linear fit.}
\label{fig2}
\end{figure}

For further treatment, we need to specify the parameters in Eqs.~(\ref{1})-(\ref{3}). We will treat the Shockley-Hall-Reed lifetime as an adjustable parameter. Let us focus on the room-temperature case. Then, at low doping levels, the recombination parameters are \cite{Sachenko06} $A_{pn} = 3.7\cdot 10^{-15}$\,cm$^{-3}$/s and $A_x = 2.6\cdot 10^{-15}$\,cm$^{-3}$/s. As the doping level increases, the electron-hole interaction becomes screened, leading to the reduction of $A_x$. To account for this effect, $A_x$ has to be multiplied by a factor $\exp\left[-0.568\left(1 - \sqrt{1 - 10^{-18}n}\right)^2\right]$, see Ref.~\onlinecite{Sachenko00-1}.

In order to find $n_x$, we use the experimental dependence of $\tau_{eff}$ on $n_0$ reported in Refs.~\onlinecite{Richter12, Fossum76, Hacker94, Beck73, Passari83}, and also the improved empirical formula for $\tau_{Auger}$ obtained in \cite{Richter12}.

The results of comparison of the experimental $\tau_{eff}(n_0)$ curve with the theoretical one, obtained using the expressions from Ref.~\onlinecite{Richter12}, are shown in Fig.~\ref{fig1}(a). The theoretical calculations were performed using Eq.~(\ref{3}), both with and without taking into account the nonradiative exciton recombination. Fig.~\ref{fig1}(b) shows the experimental dependence of $\tau_{eff}$ on the excitation level from Ref.~\onlinecite{Yablonovitch86}.

It should be noted that, in order to obtain an accurate estimate for $n_x$ from the experimental curves for $\tau_{eff}$, see Fig.~\ref{fig1}, using the relation (\ref{3}), the graphical accuracy is insufficient. Therefore, we first excluded the effect of the interband Auger recombination using the expressions from Ref.~\onlinecite{Richter12} (neglecting the radiative recombination). As a result, we obtained a simple expression
\begin{equation}
\tau_{eff}^*(n) \equiv \frac{1}{\tau_{eff}^{-1} - \tau_{Auger}^{-1} - \tau_{r1}^{-1} - \tau_{r2}^{-1}} = \frac{\tau_{SRH}}{1 + (n/n_x)}\ ,
\label{4}
\end{equation}
which implies that the combination
\begin{equation}
f(n) \equiv \frac{\tau_{SRH}}{\tau_{eff}^*(n)} - 1 = \frac{n}{n_x}
\label{5}
\end{equation}
is a linear function of $n$, whose slope is $n_x^{-1}$. Note that the function $f(n)$ is universal, i.e. independent of $\tau_{SRH}$. Fig.~\ref{fig2} shows this function, obtained theoretically and based on the experimental results of Refs.~\cite{Fossum76, Hacker94, Yablonovitch86} (with interband Auger recombination taken into account). For each of the three experimental curves, $n_x$ was found using the least squares method, yielding in all three cases very similar values 
\begin{equation}
n_x = (8.2 \pm 0.1)\cdot 10^{15}\,\text{cm}^{-3}\ .
\label{nx}
\end{equation} It is this value that was used in all theoretical curves presented in this work.

As seen from Fig.~\ref{fig1}, the noticeable discrepancy between the theoretical curves that do and do not take the nonradiative recombination into account occurs at $\tau_{SRH} = 5.2, 1.2$, and 0.4\,ms. At higher Shockley-Reed-Hall lifetimes of 30 and 40 ms, the two sets of curves are practically identical.

We note that in the work \onlinecite{Fossum76}, a value $n_x = 7.1\cdot 10^{15}$\,cm$^{-3}$ was obtained using the simplified expression (\ref{3}), where only the first term was present in the right-hand side. However, as our analysis shows, interband Auger recombination becomes operative at doping levels higher than $10^{15}$\,cm$^{-3}$, which was not taken into account in Ref.~\onlinecite{Fossum76}. Therefore, quadratic exciton nonradiative recombination plays a smaller role, and the parameter $n_x$ has a somewhat higher value.

We note that, according to (\ref{2}), the obtained value $n_x = 8.2\cdot 10^{15}$\,cm$^{-3}$ is realized when the ratio $C_t/G_t = 3.38\cdot 10^{16}$\,cm$^{-3}$. The independence of $n_x$ on the parameters of
concrete deep centers can be explained by the fact that, on the one hand, responsible for the generation-recombination processes in silicon are the levels with high capture cross-section, whose energies are close to the middle of the bandgap. On the other hand, according to Ref.~\onlinecite{Haug81}, $G_t$ depends only weakly on the specifics of the deep levels, but is determined by their energies. Therefore, for the impurity levels, whose energy is close to the middle of the bandgap, the value of $n_x$ can be practically the same.

\section{Exciton effect on the internal luminescence quantum yield in silicon}
We now apply the corrected value (\ref{nx}) to analyze the influence of the excitonic effects on the internal luminescence quantum yield in silicon, which can be written as
\begin{equation}
q = \frac{\tau_{r1}^{-1} + \tau_{r2}^{-1}}{\tau_{eff}^{-1} + (S/d)}\ ,
\label{6}
\end{equation}
where $d$ is the semiconductor slab thickness, and $S$ the total recombination velocity on its both surfaces.

For simplicity, we consider the case, where the electron-hole pair diffusion length is large: $L_{eff} = \sqrt{D\tau_{eff}} \gg d$. We will use the expression (\ref{3}) and its truncated version, in which the terms due to the radiative and nonradiative exciton recombination are omitted.

\begin{figure}[t!] 
\includegraphics[scale=0.28]{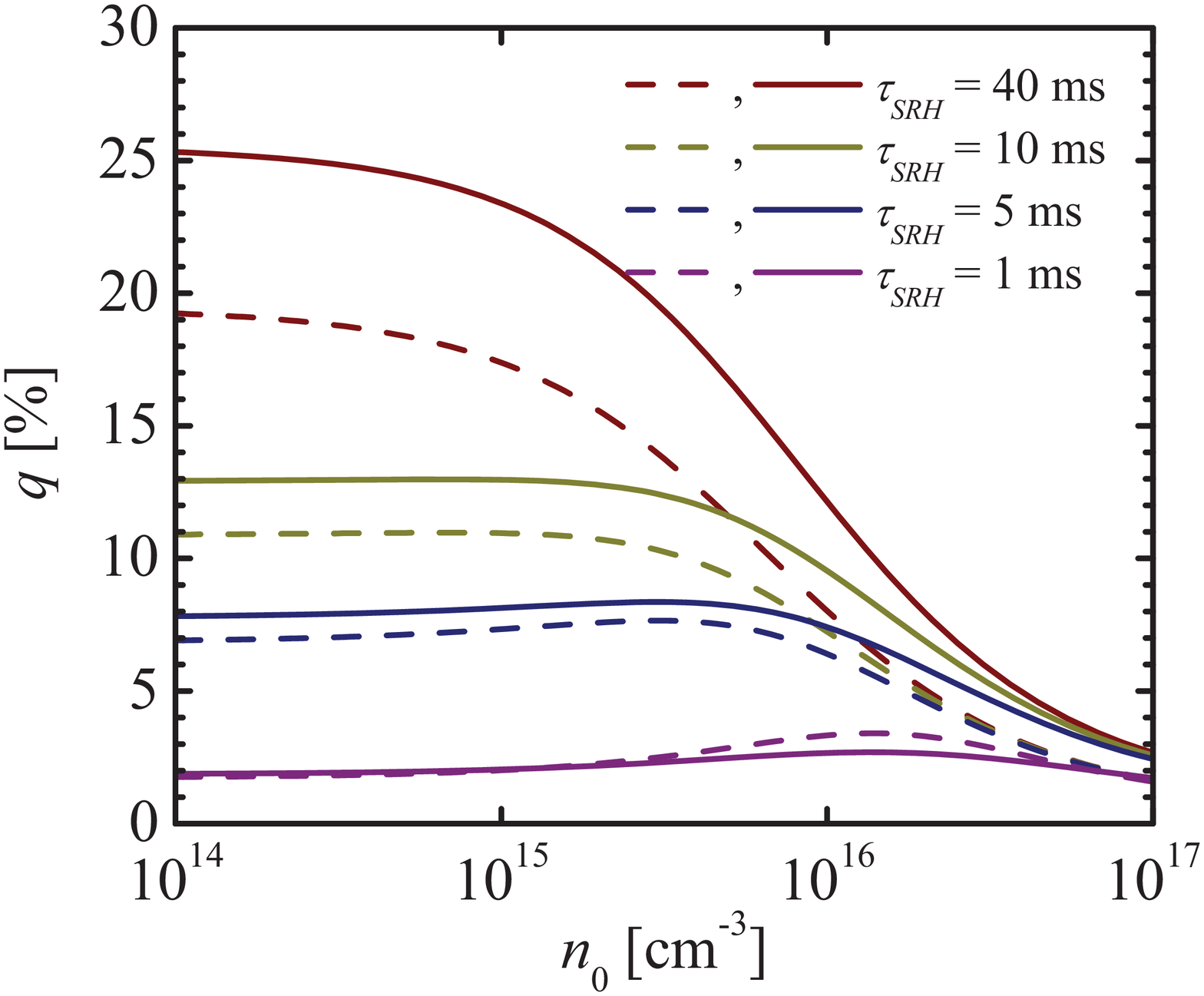}
\caption{Internal quantum yield of luminescence vs. doping level, obtained theoretically with and without exciton effects. }
\label{fig3}
\end{figure}

Fig.~\ref{fig3} shows the theoretical results for the internal luminescence quantum yield $q$ in n-type silicon. We assumed $S = 0.25$\,cm/s, the smallest value obtained in Ref.~\onlinecite{Yablonovitch86}. Furthermore, we took $\Delta n = 5\cdot 10^{15}$\,cm$^{-3}$, and $\tau_{SRH} = 40, 10, 5$, and 1\,ms. The solid curves are obtained taking the exciton effects into account, and the dashed ones are obtained without them.

As seen from Fig.~\ref{fig3}, inclusion of exciton effects leads to an increase of the luminescence quantum yield when radiative exciton recombination dominates at $\tau_{SRH} = 40, 10$, and 5\,ms. The higher $\tau_{SRH}$ the larger luminescence quantum yield obtained taking the exciton effects into account. At $\tau_{SRH} = 1$\,ms, exciton recombination practically does not affect luminescence quantum yield at $n_0 \le 10^{15}$\,cm$^{-3}$. At $n_0 > 2\cdot 10^{15}$\,cm$^{-3}$, the value of $q(n_0)$ obtained without exciton effects is larger than with them.

\begin{figure}[t!] 
\includegraphics[scale=0.28]{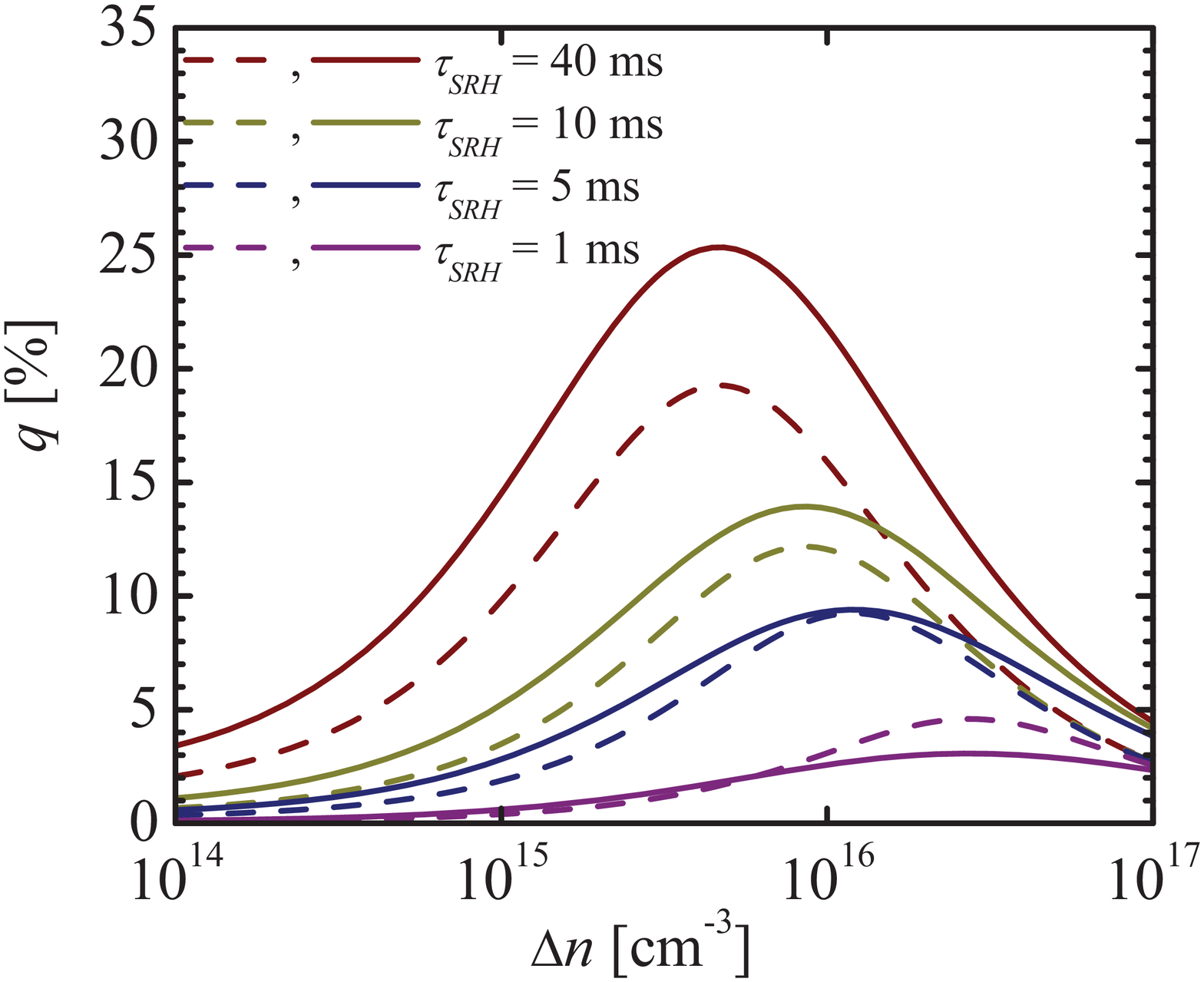}
\caption{Internal quantum yield of luminescence vs. excitation level, obtained theoretically with and without exciton effects.}
\label{fig4}
\end{figure}

Shown in Fig.~\ref{fig4} are the theoretical dependence of the luminescence quantum yield on the excitation level obtained with (solid) and without (dashed) taking the exciton effects into account. As seen from this figure, all curves have a maximum. The highest value of $q \approx 25$\,\% is realized at $\tau_{SRH} = 40$\,ms. At $\tau_{SRH} = 10$ and 40\,ms, the curves $q(\Delta n)$ obtained with exciton effects exceed the corresponding curves obtained without taking these effects into account, at all values of $\Delta n$. For $\tau_{SRH} = 5$\,ms, the curve obtained with the exciton effects taken into account exceeds the one without exciton effect almost everywhere, except for the region near the maximum. Finally, at $\tau_{SRH} = 1$\,ms, the two curves practically coincide for $\Delta n \le 10^{16}$\,cm$^{-3}$, whereas at higher $\Delta n$, the curve obtained without exciton effects is higher.

\begin{figure}[t!] 
\includegraphics[scale=0.28]{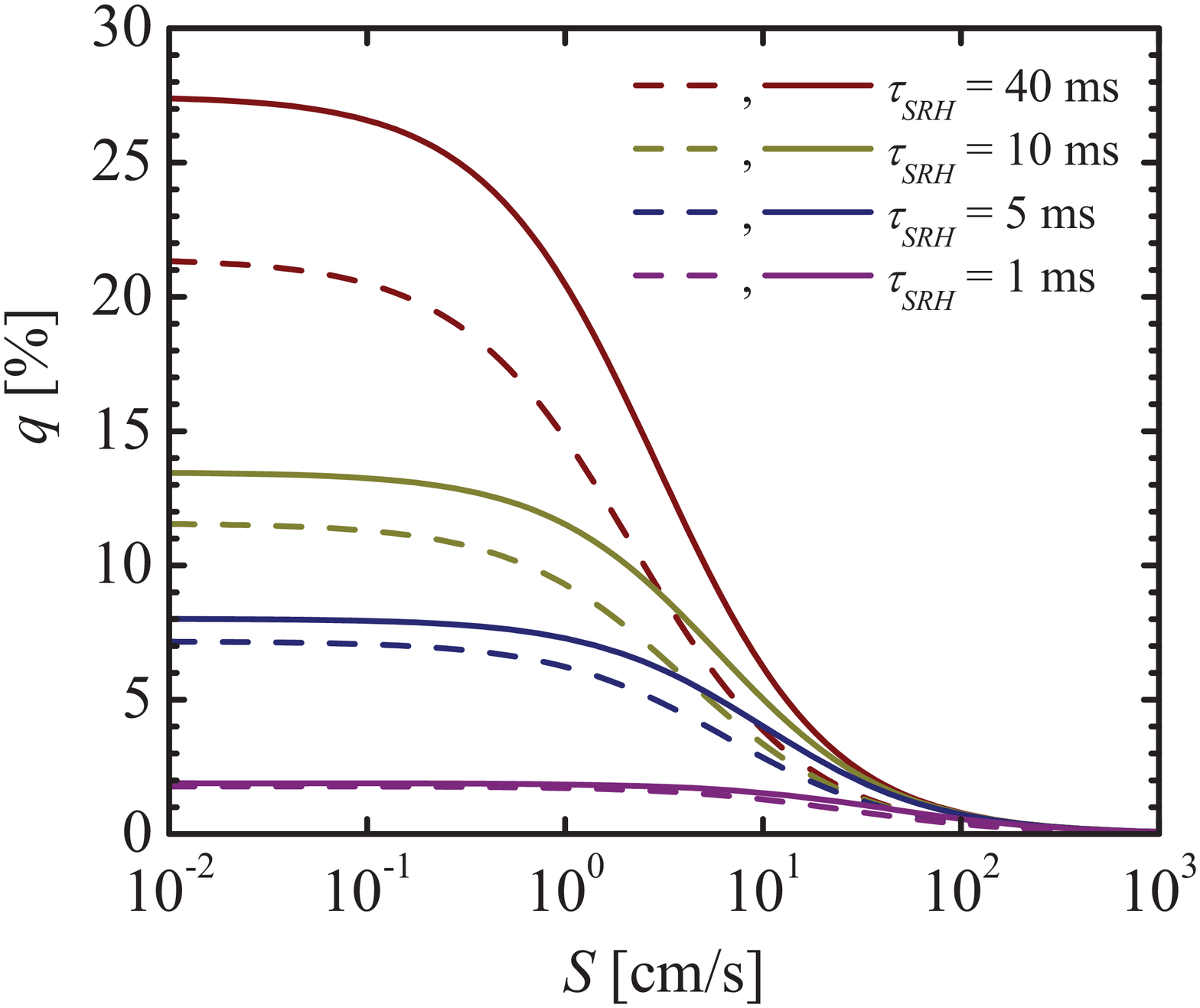}
\caption{Internal quantum yield of luminescence vs. surface recombination rate, obtained theoretically with and without exciton effects.}
\label{fig5}
\end{figure}

Presented in Fig.~\ref{fig5} is the theoretical luminescence quantum yield as a function of surface recombination velocity $S$ with and without exciton effects (solid and dashed curves, respectively). As seen in this plot, exciton effects make $q(S)$ higher at $\tau_{SRH} = 40, 10$, and 5\,ms; the two curves for $\tau_{SRH} = 1$\,ms practically coincide. The higher $\tau_{SRH}$, the smaller the surface recombination velocity, at which $q(S)$ starts to decrease. At the typical value of $S$ of the order of $10^3$\,cm/s, quantum yield does not exceed 0.1\,\%. Thus, in order to increase luminescence quantum yield, efficient methods of surface passivation are required, see Refs.~\onlinecite{Richter12, Yablonovitch86} for a description of such methods. In particular, it is possible to substantially reduce surface recombination velocity by using $\alpha$-Si:H layers of nanometer thickness \cite{Jano13}.

We note that the curves from Figs.~\ref{fig3}-\ref{fig5} are obtained under the assumption that the doping level, $n_0$, and the excitation level, $\Delta n$, are independent parameters. The latter is proportional to the irradiation intensity; however, it also depends on the recombination mechanisms, including the interband Auger recombination. Therefore $n_0$ and $\Delta n$ are, in fact, related. This relation can be found from the generation-recombination balance equation for the photodiode silicon structures used to investigate electroluminescensce. In the open-circuit regime, and for $L_{eff} \gg d$, this equation has the form:
\begin{equation}
I_{SC}/q = A_{SC}\left((d/\tau_{eff}) + S\right)\Delta n\ ,
\label{7}
\end{equation}
where $I_{SC}$ is the short-circuit current, and $A_{SC}$ is the area of the semiconductor structure.

\begin{figure}[t!] 
\includegraphics[scale=0.28]{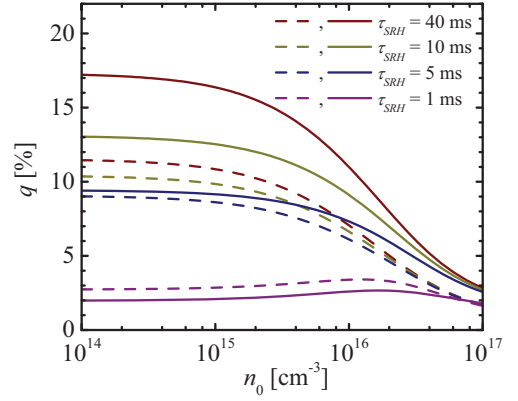}
\caption{Internal quantum yield of luminescence vs. doping level, obtained theoretically with and without exciton effects.}
\label{fig6}
\end{figure}

Fig.~\ref{fig6} shows the internal luminescence quantum yield as a function of doping level for AM1.5 conditions. As before, these curves are build with and without taking exciton effects into account (solid and dashed lines, respectively). The photocurrent density was assumed to be $J_{SC} = I_{SC}/A_{SC} = 39.5$\,mA/cm$^2$. To build these curves, we first determined $\Delta n(n_0)$ from Eq.~(\ref{7}). Then, this dependence was substituted into Eq.~(\ref{6}) and into the reduced version of Eq.~(\ref{6}), where the recombination terms due to radiative and nonradiative exciton recombination are omitted.

As seen from Fig.~\ref{fig6}, the curves obtained are very similar to the ones from Fig.~\ref{fig3}. However, there is a difference between the two dependences. The main difference is that the magnitude of internal yield in Fig.~\ref{fig6} is smaller than in Fig.~\ref{fig3}. This is due to the fact that the initial $\Delta n$ obtained from Eq.~(\ref{5}) at sufficiently small $n_0$ are higher than $10^{16}$\,cm$^{-3}$. At the same time, as seen from Fig.~\ref{fig4}, the maximum of the $q(\Delta n)$ curve at $\tau_{SRH} = 40, 10$, and 5\,ms is below $10^{16}$\,cm$^{-3}$. This means that, in order to increase $q(\Delta n)$, one needs to reduce $\Delta n$, which is achieved by the reduction of the photogenerated current, i.e. by the reduction of the irradiation intensity. Indeed, for $J_{SC} \approx 4$\,mA/cm$^2$ we obtain $q \approx 25$\,\% at $\tau_{SRH} = 40$\,ms. This agrees with the result from Fig.~\ref{fig3}.

\section{Conclusions}
As shown in this work, the effective density $n_x$, which determines the nonradiative exciton lifetime in silicon is $8.2\cdot 10^{15}$\,cm$^{-3}$. Our analysis has revealed that the exciton effects in silicon lead to an increase of the internal luminescence quantum yield at sufficiently long Shockley-Reed-Hall lifetimes exceeding 1\,ms. In the opposite case, $\tau_{SRH} < 1$\,ms, this effect is absent.

\section*{Acknowledgments}
 M.E. is grateful to the Natural Sciences and Engineering Research Council of Canada (NSERC) and to the Research and Development Corporation of Newfoundland and Labrador (RDC) for financial support.

\end{document}